\documentclass[conference]{IEEEtran}
\IEEEoverridecommandlockouts
\usepackage{cite}
\usepackage[numbers]{natbib}
\usepackage{amsmath,amssymb,amsfonts}
\usepackage{algorithmic}
\usepackage{graphicx}
\usepackage{textcomp}
\usepackage{xcolor}
\usepackage{url}
\def\BibTeX{{\rm B\kern-.05em{\sc i\kern-.025em b}\kern-.08em
 T\kern-.1667em\lower.7ex\hbox{E}\kern-.125emX}}
\begin{document}

\title{GiveMeLabeledIssues: An Open Source Issue Recommendation System\\
}

\author{\IEEEauthorblockN{Joseph Vargovich, Fabio Santos, Jacob Penney, Marco A. Gerosa, Igor Steinmacher}
\IEEEauthorblockA{\textit{School of Informatics, Computing, and Cyber Systems} \\
\textit{Northern Arizona University}\\
Flagstaff, United States \\
joseph\_vargovich@nau.edu, fabio\_santos@nau.edu, jacob\_penney@nau.edu, Marco.Gerosa@nau.edu, Igor.Steinmacher@nau.edu}
}

\maketitle

\begin{abstract}

Developers often struggle to navigate an Open Source Software (OSS) project's issue-tracking system and find a suitable task. Proper issue labeling can aid task selection, but current tools are limited to classifying the issues according to their type (e.g., bug, question, good first issue, feature, etc.). In contrast, this paper presents a tool (\textit{GiveMeLabeledIssues}) that mines project repositories and labels issues based on the skills required to solve them. We leverage the domain of the APIs involved in the solution (e.g., User Interface (UI), Test, Databases (DB), etc.) as a proxy for the required skills. \textit{GiveMeLabeledIssues} facilitates matching developers' skills to tasks, reducing the burden on project maintainers. The tool obtained a precision of 83.9\% when predicting the API domains involved in the issues. The replication package contains instructions on executing the tool and including new projects. A demo video is available at https://www.youtube.com/watch?v=ic2quUue7i8 


\end{abstract}


\section{Introduction}
Contributors struggle to find issues to work on in Open Source Software (OSS) projects due to difficulty determining the skills required to work on an issue \cite{steinmacher2015systematic}. Labeling issues manually can help \cite{santos2022choose}, but manual work increases maintainers' effort, and many projects end up not labeling their issues~\cite{barcomb2020managing}. 


To facilitate issue labeling, several studies proposed mining software repositories to tag the issues with labels such as bug/non-bug and good-first-issue (to help newcomers find a suitable task) \cite{pingclasai2013classifying,zhu2019bug, el2020automatic, perez2021bug}. However, newcomers to a project have a variety of skill levels and these approaches do not indicate the skills needed to work on the tasks. 

APIs encapsulate modules with specific functionality. If we can predict APIs used to solve an issue, we can guide new contributors on what to contribute. However, since projects include thousands of different APIs, adding each API as a label would harm the user experience. To keep the number of labels manageable, we classified the APIs into domains for labeling~\cite{santos2023tell,santos2022choose}

In our study, API domains represent categories of APIs such as ``UI,'' ``DB,'' ``Test,'' etc. In our previous work, we used 31 distinct API domains \cite{santos2023tell}. However, we acknowledge that the set of possible API-domain labels varies by project, as each project requires different skills and expertise. 
In another work, Santos et al.~\cite{santos2021can} have evaluated the API domain labels in an empirical experiment with developers. The study showed that the labels enabled contributors to find OSS tasks that matched their expertise more efficiently. 
Santos et al. study showed promising results when classifying issues by API-Domain, with precision, recall, and F-measure scores of 0.755, 0.747, and 0.751, respectively. 

Following this idea, we implemented \textit{GiveMeLabeledIssues} to classify issues for potential contributors. \textit{GiveMeLabeledIssues} is a web tool that indicates API-domain labels for open issues in registered projects. Currently, \textit{GiveMeLabeledIssues} works with three open-source projects: JabRef, PowerToys, and Audacity. The tool enables users to select a project, input their skill set (in terms of our categories), and receive a list of open issues that would require those skills. 

We trained the tool with the title and body text of closed issues and the APIs utilized within code that solve the issue, as gathered from pull requests. 
We evaluated the tool using closed issues as the ground truth and found an average of 83.8\% precision and 78.5\% recall when training and testing models for individual projects. 




\section{GiveMeLabeledIssues Architecture}\label{ToolDesc}

\textit{GiveMeLabeledIssues} leverages prediction models trained with closed issues linked with merged pull requests. On top of these models, we built a platform that receives user input and queries the open issues based on the users' skills and desired projects. In the following, we detail the process of building the model, the issue classification process, and the user interface that enables users to get the recommendation. \textit{GiveMeLabeledIssues} is structured in two layers: the frontend web interface\footnote{https://github.com/JoeyV55/GiveMeLabeledIssuesUI} and the backend REST API\footnote{https://github.com/JoeyV55/GiveMeLabeledIssuesAPI}.

\subsection{Model training}

\begin{figure}[!hbt]
 \centering
 \includegraphics[width=.5\textwidth] {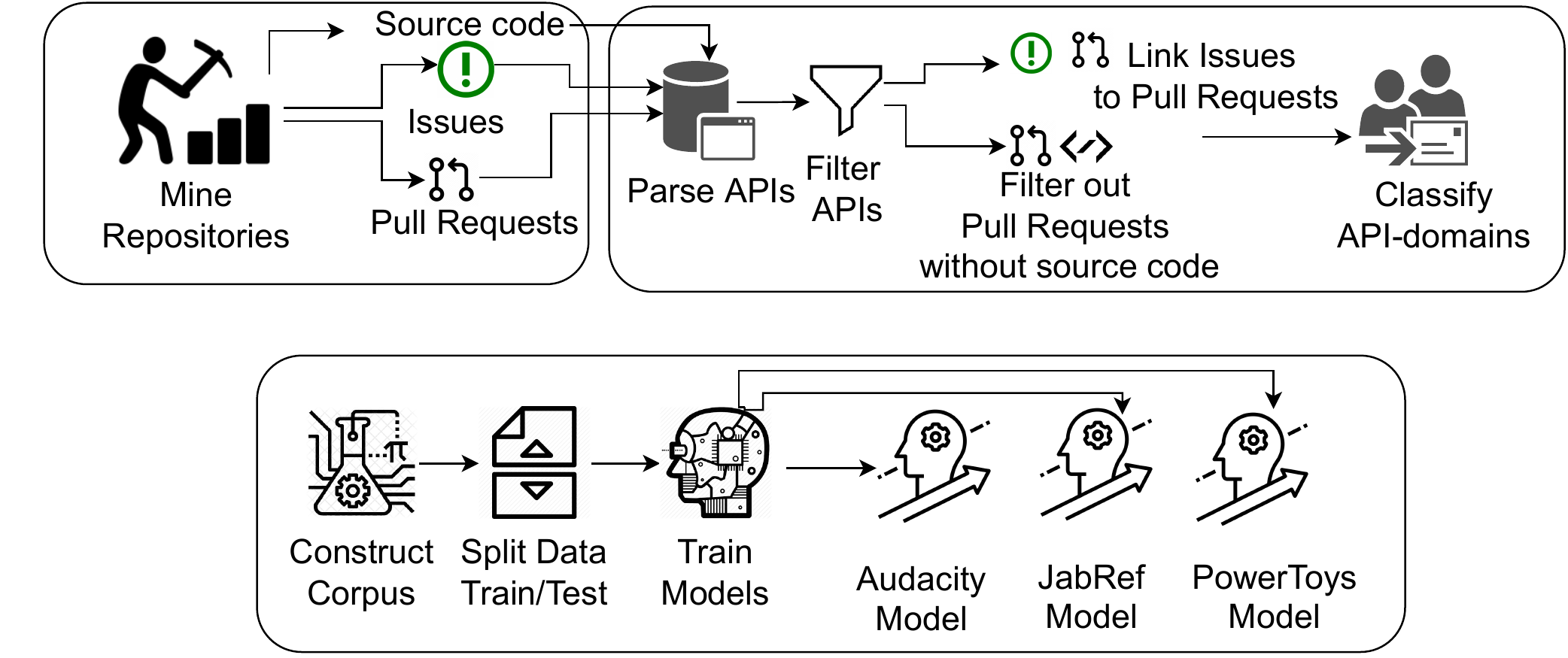}
 \caption{The Process of Training a Model}
 \label{fig:training}
\end{figure}

In the current implementation, a model must be built for each project using data collected from project issues linked with merged pull requests. The tool maps the issue text data to the APIs used in the source code that solved the issues (Figure \ref{fig:training}). 

\subsubsection{Mining repositories}

We collected 18,482 issues and 3,129 pull requests (PRs) from JabRef, PowerToys, and Audacity projects up to November 2021. We used the GitHub REST API v3 to collect the title, body, comments, closure date, name of the files changed in the PR, and commit messages.

\subsubsection{APIs parsing}
We built a parser to process all source files from the projects to identify the APIs used in the source code affected by each pull request. In total, we found 3,686 different APIs in 3,108 source files. The parser looked for specific commands, i.e., import (Java), using (C\#), and include (C++). The parser identified all classes, including the complete namespace from each import/using/include statement.

\subsubsection{Dataset construction}
We kept only the data from issues linked with merged and closed pull requests since we needed to map issue data to the APIs that are used in the source code files changed to close the issue. To find the links between pull requests and issues, we searched for the symbol \texttt{\#issue\_number} in the pull request title and body and checked the URL associated with each link. We also filtered out issues linked to pull requests without at least one source code file (e.g., those associated only with documentation files) since they do not provide the model with content related to any API.

\subsubsection{API categorization}
We use the API domain categories defined by \citet{santos2023tell}. These 31 categories were defined by experts to encompass APIs from several projects (e.g., UI, IO, Cloud, DB, etc.---see our replication package\footnote{https://doi.org/10.5281/zenodo.7575116}. 


\subsubsection{Corpus construction}
We used the issue title and body as our corpus to train our model since they performed well in our previous analysis~\cite{santos2021can}. Similar to other studies~\cite{behl2014bug,vadlamani2020studying}, we applied TF-IDF as a technique for quantifying word importance in documents by assigning a weight to each word following the same process described in the previous work \cite{santos2021can}. TF-IDF returns a vector whose length is the number of terms used to calculate the scores. Before calculating the scores, we convert each word to lowercase and removed URLs, source code, numbers, and punctuation. After that, we remove templates and stop-words and stem the words. These TF-IDF scores are then passed to the Random Forest classifier (RF) as features for prediction. RF was chosen since it obtained the best results in previous work \cite{santos2021can}. The ground truth has a binary value (0 or 1) for each API domain, indicating whether the corresponding domain is present in the issue solution.

We also offer the option of using a BERT model in \textit{GiveMeLabeledIssues}. We created two separate CSV files to train BERT: an input binary with expert API-domain labels paired with the issue corpus and a list of the possible labels for the specific project. BERT directly labels the issue using the corpus text and lists possible labels without needing an additional classifier (such as Random Forest). 

\subsubsection{Building the model}
The BERT model was built using the Python package fast-bert\footnote{https://github.com/utterworks/fast-bert}, which builds on the Transformers\footnote{\url{https://huggingface.co/docs/transformers/index}} library for Pytorch. Before training the model, the optimal learning rate was computed using a LAMB optimizer~\cite{You2020Large}. Finally, the model was fit over 11 epochs and validated every epoch. The BERT model was trained on an NVIDIA Tesla V100 GPU with an Intel(R) Xeon(R) Gold 6132 CPU within a computing cluster.

TF-IDF and BERT models were trained and validated for every fold in a ShuffleSplit 10-fold cross-validation. Once trained, the models were hosted on the backend. The replication package contains instructions on registering a new project by running the model training pipeline that feeds the demo tool. The models can then output predictions quickly without continually retraining with each request. 

For the RandomForestClassifier (TF-IDF), the best classifier, we kept the following parameters: criterion = 'entropy,' max depth = 50, min samples leaf = 1, min samples split=3, and n estimators = 50.

\subsection{Issue Classification Process}

\textit{GiveMeLabeledIssues} classifies currently open issues for each registered project. The tool combines the title and body text and sends it to the classifier. The classifier then determines which domain labels are relevant to the gathered issues based on the inputted issue text. The labeled issues are stored in an SQLite database for future requests, recording the issue number, title, body, and domain labels outputted by the classifier.

\begin{figure}[!hbt]
 \centering
 \includegraphics[width=.5\textwidth] {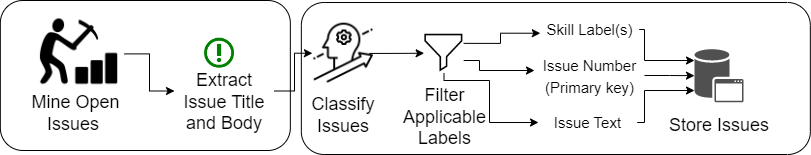}
 \caption{The Process of Classifying and Storing Issues}
 \label{fig:Classify}
\end{figure}

The open issues for all projects registered with \textit{GiveMeLabeledIssues} are reclassified daily to ensure that the database is up to date as issues are opened and closed. Figure \ref{fig:Classify} outlines the daily classification procedure. 

\subsection{User Interface}

\textit{GiveMeLabeledIssues} outputs the labeled issues to the User Interface. The user interface is implemented using the Angular web framework. To use the tool, users provide the project name and select API-domain labels they are interested in. 
This information is sent to the backend REST endpoint via a GET request. The backend processes the request, recommending a set of relevant issues for the user. 

\begin{figure}[!h]
 \centering
 \includegraphics[width=.5\textwidth] {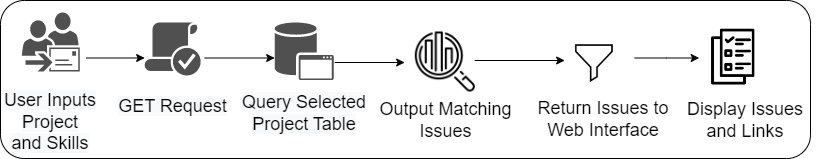}
 \caption{The Process of Outputting Issues}
 \label{fig:OutputtingSteps}
 \end{figure}

The backend REST API is implemented using the Django Rest Framework. It houses the trained TF-IDF and BERT text classification models and provides an interface to the labeled issues. When receiving the request, the backend queries the selected project for issues that match the user's skills. Once the query is completed, the backend returns the labeled issues to the user interface. Each labeled issue includes a link to the open issue page on GitHub and the issue's title, number, and applicable labels. The querying process is shown in Figure \ref{fig:OutputtingSteps}.

\begin{figure}[!hbt]
 \includegraphics [width=.5\textwidth] {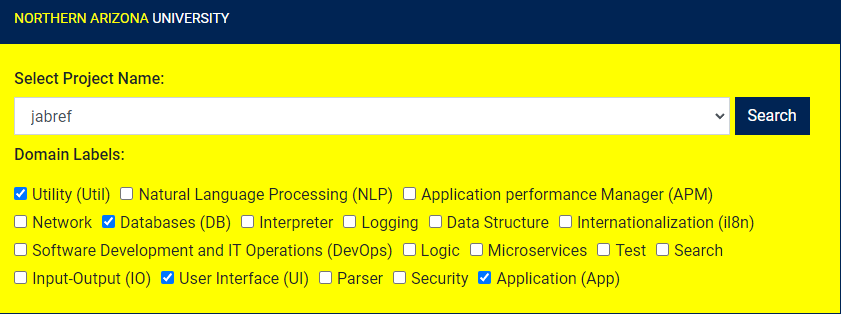}
 \caption{Selection of a project and API domains}
 \label{fig:skillSelect}
 \end{figure}

\begin{figure}[!hbt]
 \includegraphics [width=.5\textwidth] {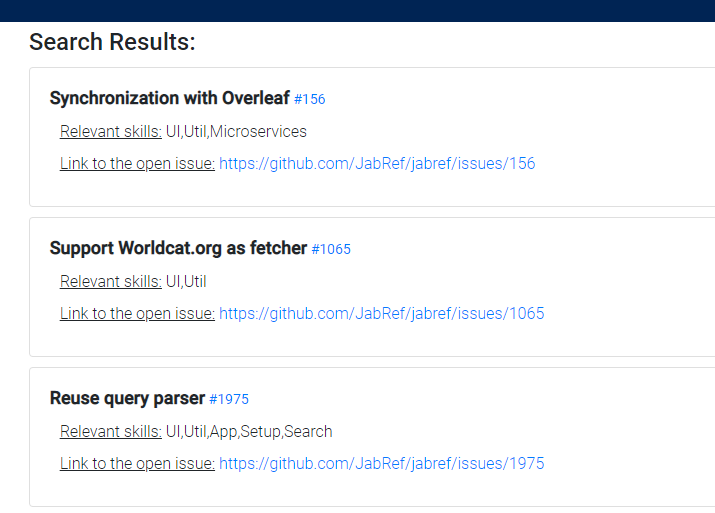}
 \caption{Labeled Issues Outputted for JabRef with the Utility, Databases, User Interface, and Application skills Selected}
 \label{fig:issueResult}
 \end{figure}

Figure~\ref{fig:skillSelect} shows JabRef selected as the project and ``Utility,'' ``Databases,'' ``User Interface,'' and ``Application'' as the API domains provided by the user. Figure~\ref{fig:issueResult} shows the results of this query, which displays all JabRef open issues that match those labels.

\section{Evaluation}

We have evaluated the performance of the models used to output API-domain labels using a dataset comprised of 
18,482 issues, 3,129 PRs, 3,108 source code files, and 3,686 distinct APIs, chosen from active projects and diverse domains: Audacity (C++), PowerToys (C\#), and JabRef (Java).\footnote{The model training replication package is available at \url{https://zenodo.org/record/7726323#.ZA5oy-zMIeY}} 
Audacity is an audio editor, PowerToys is a set of utilities for Windows, and JabRef is an open-source bibliography manager. 
Table \ref{tab:resultsTFIDF-BERT} shows the results with Precision, Recall, F-Measure, and Hamming Loss values. 
We trained models to predict API-domain labels using individual issue datasets from each project and a single dataset that combined the data from all the projects. 


\begin{table}[]
\centering
 \caption{Overall metrics from models -averages. RF-Random Forest* Hla - Hamming Loss**}
 \label{tab:resultsTFIDF-BERT}
\begin{tabular}{|c|c|c|r|r|r|}
\hline
\begin{tabular}[c]{@{}c@{}}Model \\ averages\end{tabular} & \begin{tabular}[c]{@{}c@{}}\textbf{O}ne/\textbf{M}ulti \\ projects\end{tabular} & Precision & Recall & \begin{tabular}[c]{@{}c@{}}F-meas \\-ure \end{tabular}& Hla** \\ \hline
RF* TF-IDF & O & \textbf{0.839} & 0.799 & 0.817 & \textbf{0.113} \\
BERT & O & 0.595 & 0.559 & 0.568 & 0.269 \\
RF* TF-IDF & M & 0.659 & 0.785 & 0.573 & 0.153 \\
BERT & M & 0.593 & 0.725 & 0.511 & 0.219 \\ 
\citet{izadi2022predicting} & M & 0.796 & 0.776 & 0.766 & NA \\
\citet{kallis2019ticket} & M & 0.832 & \textbf{0.826} & \textbf{0.826} & NA \\
\citet{santos2021can} & O & 0.755 & 0.747 &
0.751 & 0.142 \\
\hline
\end{tabular}
\end{table}

\begin{table}[]
\centering
 \caption{Overall metrics from models - by projects. RF-Random Forest* Hla - Hamming Loss**}
 \label{tab:resultsTFIDF-BERT-projects}
\begin{tabular}{|c|c|c|r|r|r|}
\hline
\begin{tabular}[c]{@{}c@{}}Model \\ by project\end{tabular} & \begin{tabular}[c]{@{}c@{}}Project\end{tabular} & Precision & Recall & \begin{tabular}[c]{@{}c@{}}F-meas \\-ure \end{tabular}& Hla** \\ \hline
RF* TF-IDF & Audacity	&\textbf{0.872}	&\textbf{0.839}	&\textbf{0.854} & 0.103\\	
RF* TF-IDF & JabRef	&0.806	&0.782	&0.793 & 0.143\\	
RF* TF-IDF & PowerToys	&0.84	&0.776	&0.805 &\textbf{0.094}\\	
BERT & Audacity		&0.382	&0.511	&0.434 & 0.42\\	
BERT & JabRef	&0.791	&0.606	&0.686	& 0.192\\
BERT & PowerToys	&0.619	&0.643	&0.626 & 0.187\\	
\hline
\end{tabular}
\end{table}

As shown in Table~\ref{tab:resultsTFIDF-BERT}, TF-IDF overcame BERT both in the per-project analysis and for the complete dataset. The difference was quite large when using single projects. The results were closer when we used the combined dataset that included data from all projects (3,736 linked issues and pull requests). We hypothesize that the sample size influenced the classifiers' performance. This aligns with previous research on issue labeling that showed that BERT performed better than other language models for datasets larger than 5,000 issues \cite{wang2021well}. TF-IDF performs very well when the dataset is from a single project because the vocabulary used in the project is very contextual, and the frequency of terms can identify different aspects of each issue. When we include the dataset from all the projects, the performance of TF-IDF drops as the context is not unique. These results outperformed the results from the API-domain labels case study conducted by Santos et al. \cite{santos2021can}. 
The project metrics (Table \ref{tab:resultsTFIDF-BERT-projects}) varied less than 6\% (e.g., the recall: 0.839 (Audacity) - 0.776 (PowerToys). Audacity had the best scores for all metrics except Hamming Loss. 

\section{Related Work}
The existing literature explores strategies to help newcomers find tasks within OSS projects. Steinmacher et al.~\cite{steinmacher2016overcoming} proposed a portal to help newcomers find relevant documentation. Despite pointing the contributor to existing resources, a newcomer may have difficulty relating the documentation to the skills required to solve a task. 

There are also several approaches designed to label issues automatically. However, most of them only try to distinguish bug reports from non-bug reports \cite{pingclasai2013classifying,zhou2016combining,xia2013tag}. \citet{zhou2016combining} built a Naive Bayes (NB) classifier to predict the priorities of bug reports. \citet{xia2013tag} tagged questions using preexisting labels. Other work \cite{el2020automatic, perez2021bug} is also restricted to existing labels while others \cite{kallis2019ticket, izadi2022predicting} proposed other labels. \citet{kallis2019ticket}, for instance, employed the textual description of the issues to classify the issues into types.

Labeling also attracted the attention of software requirement researchers. 
\citet{perez2021requirements} and \citet{quba2021software} proposed to categorize documents with functional and non-functional requirements labels to solve software engineers' problems. Non-functional requirements labels included 12 general domains like ``Performance'' or ``Availability''
\cite{quba2021software}. Such approaches use higher-level labels that would not guide contributors to choose appropriate tasks given their skills. ``Availability,'' for instance, may be related to a database, network, or a cloud module and can be challenging for a newcomer to realize where to start due to the extent of modules to analyze to find the root cause of a bug. 

APIs are also often investigated in software engineering. Recent work focuses on providing API recommendation \cite{8186224, 8478004}, giving API tips to developers \cite{8816774}, defining the skill space for APIs, developers, and projects \cite{dey2020representation} or identifying experts in libraries \cite{8816776}. The APIs usually implement a concept (database access, thread management, etc.). Knowing the API involved in a potential solution allows newcomers to pick an issue that matches their skillset. Therefore, unlike the presented related studies, our tool labels issues based on API domains~\cite{santos2021can}. 

\section{Conclusion and Future Work}
\label{conclusion}

\textit{GiveMeLabeledIssues} provides OSS developers with issues that can potentially match their skill sets or interests. Such a tool may facilitate the onboarding of newcomers and alleviate the workload of project maintainers.  

Future work can explore different domain labels, such as those offered by accredited standards (i.e., ACM/IEEE/ISO/SWEBOK). 
As a future step to evaluate the tool's impact, we will conduct a study to receive feedback from contributors and assess how the tool influences their choices by means of controlled experiments. Future work can also incorporate the use of social features~\cite{santos2023tell} and integrate the tool into GitHub Actions, or bots \cite{kinsman2021software}.

\section*{Acknowledgments}

This work is partially supported by the National Science Foundation under Grant numbers 1815503, 1900903, and 2236198.

\bibliography{demo}

\end{document}